\documentclass[aps,prl,twocolumn,longbibliography,notitlepage,superscriptaddress,preprintnumbers,nobibnotes,nofootinbib]{revtex4-1}
\usepackage{graphicx}
\usepackage{amsmath, amsfonts, amssymb}
\usepackage{mathrsfs}
\usepackage{hyperref}
\hypersetup{
    pdfstartview={FitH},
    pdfpagemode={None}
    }
\usepackage{bm}

\providecommand\figwidth{3.375in}

\graphicspath{{Figs/}}

\providecommand{\abs}[1]{\left\lvert#1\right\rvert}   
\providecommand{\av}[1]{\left<#1\right>}
\providecommand{\ket}[1]{\left|#1\right\rangle}

\providecommand{\bra}[1]{\left\langle#1\right|}

\providecommand{\del}{\partial}

\providecommand{\DE}{\Delta E}
\providecommand{\DEp}{\Delta E^\prime}
\providecommand{\dQ}{\Delta Q} 
\providecommand{\HIB}{h}

\providecommand{\ehi}{\varepsilon_1}
\providecommand{\elo}{\varepsilon_2}
\providecommand{\ebare}{\varepsilon_0}

\begin{document}
\title{Nonradiative lifetimes in intermediate band materials -- absence of lifetime recovery}
\author{Jacob J.\ Krich}
\affiliation{Department of Chemistry and Chemical Biology, Harvard University, Cambridge, MA 02138}
\author{Bertrand I.\ Halperin}
\affiliation{Department of Physics, Harvard University, Cambridge, MA 02138}
\author{Al\'an Aspuru-Guzik}
\affiliation{Department of Chemistry and Chemical Biology, Harvard University, Cambridge, MA 02138}
\date{\today}

\begin{abstract}
Intermediate band photovoltaics hold the promise of being highly efficient and cost effective photovoltaic cells. Intermediate states in the band gap, however, are known to facilitate nonradiative recombination. Much effort has been dedicated to producing metallic intermediate bands in hopes of producing \emph{lifetime recovery} -- an increase in carrier lifetime as doping levels increase. We show that lifetime recovery induced by the insulator-to-metal transition will not occur, because the metallic extended states will be localised by phonons during the recombination process. Only trivial forms of lifetime recovery, e.g., from an overall shift in intermediate levels, are possible. Future work in intermediate band photovoltaics must focus on optimizing subgap optical absorption and minimizing recombination, but not via lifetime recovery.
\end{abstract}
\maketitle

The development of novel highly-efficient photovoltaic (PV) devices has the potential to significantly address the global energy and carbon problems. 
The vast majority of commercial solar cells are made from single-junction semiconductors, an architecture which Shockley and Queisser showed has an absolute efficiency limit of 41\% (with concentrated sunlight) \cite{Shockley61}. Among the proposals to break this limit is the intermediate band (IB) photovoltaic, which has an efficiency limit of 63\%, considerably higher than the single-junction limit \cite{Luque97,Luque10}.

\begin{figure}
	\includegraphics[width=\figwidth]{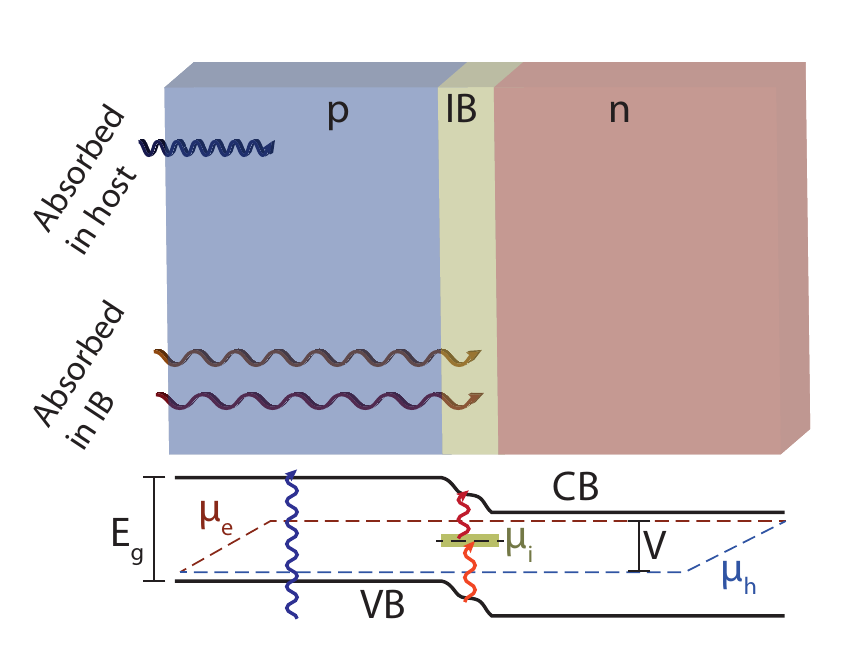}
	\caption{\label{fig:IBdevice} An intermediate band (IB) device has an IB layer between standard p- and n-type semiconductor layers. The IB layer absorbs subgap photons, which pass through the p-type (or n-type) host layer, increasing current generation. At bottom, band levels under illumination are shown, indicating the voltage $V$ and the three separate quasi-Fermi levels $\mu_e$, $\mu_h$, and $\mu_i$ for the CB electrons, VB holes, and IB states, respectively. If trapping rates are low, voltage production is still determined by the n- and p-type band gaps.}
\end{figure}

A standard single-junction semiconductor PV must optimise its band gap to maximise the product of current and voltage; these compete because it can absorb only photons with energy greater than the band gap $E_g$, and the supplied voltage can be no larger than $E_g/e$, where $-e$ is the electron charge. An IBPV device, illustrated in Fig.\ \ref{fig:IBdevice}, has an extra set of levels inside the semiconductor band gap; two subgap photons can be absorbed by the IB layer, producing a single electron-hole pair. Electrical contact is made only to the standard n- and p-type layers, so the IB layer produces extra current while allowing the full band gap to set the limit on the voltage, giving the considerably elevated efficiency bound of 63\% \cite{Luque97}. The IBPV effect has been demonstrated in a number of systems \cite{Lopez11,Wang09,Marti06}, though it has not yet produced high efficiency cells.

One method for making a material with an IB is to dope a semiconductor with large concentrations of dopants that form donor (or acceptor) levels deep inside the band gap. There is an obvious problem with this recipe: levels deep in the band gap are well known to cause nonradiative recombination \cite{Shockley52,Hall52}. It was proposed that a sufficiently high concentration of dopants could cause an insulator to metal transition (IMT) in the IB, which would suppress the nonradiative recombination rate and cause \emph{lifetime recovery} in which adding additional dopants decreases the nonradiative recombination rate \cite{Luque06}. A great deal of work has gone into looking for such an IMT in doped semiconductors \cite{Olea09,Olea10,Sanchez10,Krich11,Winkler11,Olea11a}, including a report of lifetime recovery \cite{Antolin09}.

The theoretical claim of lifetime recovery arose from a study of multiphonon recombination \cite{Huang50,Lang75,Henry77}. A single defect-mediated recombination event consists of two trapping events: a conduction band (CB) electron is trapped by an unoccupied defect state and a valence band (VB) hole is trapped by an occupied defect state. The statistical mechanics of the occupancy of the trap levels -- showing that midgap states are best at fostering nonradiative recombination -- was worked out by Shockley and Read \cite{Shockley52} and Hall \cite{Hall52} and is called SRH recombination. The theory of multiphonon trapping is well developed for isolated impurities \cite{Henry77,Huang50,Landsberg91}. When applied to polar semiconductors, that theory implies that the trapping rates decrease as the size of the defect wavefunction increases \cite{Landsberg91,Malm71}; this result was taken to imply that midgap metallic, extended states would not induce recombination \cite{Luque06}.

Here, we extend the multiphonon recombination theory to the case of many impurities. 
We show that delocalised states actually increase trapping rates compared to an equivalent localised system. We show that as long as the shift of IB energies due to interactions in the IB is less than the dopant ionization energy $\DEp$, the trapping rates cannot be significantly reduced from the independent-dopant limit, regardless of whether the IB eigenstates are extended or localised. A trivial form of lifetime recovery in which the interaction between defects moves the IB away from mid-gap is possible, but this is not a useful route to IBPV, as the optical absorption frequencies will change accordingly. A similarly trivial form of lifetime recovery could also occur if the dopant chemical state changes with doping, e.g., if precipitates form. 

Despite this negative prediction about true lifetime recovery, IBPV still has potential to produce highly efficient solar cells if materials can be found which have sufficiently strong absorptivity for subgap photons, as we discuss at the end of this article. Efforts should be directed toward finding semiconductor:dopant systems which have intrinsically long trapping times and in which the subgap absorptivities are high, not toward finding insulator-to-metal transitions in doped semiconductors.

\section{Trapping rate}
\begin{figure}
  \includegraphics[width=\figwidth]{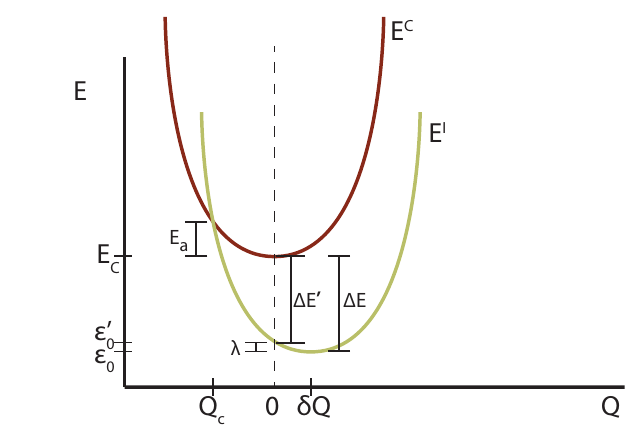}
  \caption{\label{fig:isolated} Energy levels as a function of lattice distortion $Q$ with the electron in the CB (upper) and with the electron in the impurity state (lower). The equilibrium phonon coordinate is shifted when the localised impurity state is occupied, giving the relaxation energy $\lambda$ and causing the parabolas to cross. The activation energy $E_a$ shows the energy required to take the CB electron to the degeneracy point, from which trapping occurs most rapidly.}
\end{figure}

The argument for lifetime recovery focuses on multiphonon recombination, and we will consider that mechanism here. For neutral defects and deep defect levels, the multiphonon trapping mechanism is believed to be most important; in other cases cascade capture is important \cite{Lax60}. 
The physics inside an IB can be complicated, with both disorder and interactions. The band edges of the IB, however, should be relatively easy to determine optically, so we consider the energies $\ehi$ and $\elo$ of the highest and lowest energy states in the IB to be known. 
In the standard multiphonon trapping process, we consider a conduction band (CB) electron (equivalently a valence band hole, but we will consider an electron for specificity) captured by an unoccupied defect state (see Fig.\ \ref{fig:isolated}).

We work in the Born-Oppenheimer approximation, in which the lattice is considered static for the electronic system, so we neglect the phonon momentum; each phonon coordinate $Q_i$ has an associated angular frequency $\omega_i$, and there are $N$ impurities. We consider a situation in which there is initially one electron in the CB in a state $\ket{\phi_c}$, and all IB states are empty. This situation permits a single-electron picture with $N$ localised orthogonalised impurity wavefunctions $\ket{\psi_\alpha}$ for $\alpha=1\dots N$. We discuss effects of partial filling of the IB at the end of this paper. In this approximation, we ignore scattering by CB states into each other, which is valid in the usual case that the CB bandwidth is large compared to the electron-phonon coupling. Assuming linear coupling between the electron and phonon degrees of freedom, the Hamiltonian can be written as $H=H_{el}+E_{ph}+H_{el-ph}$ with
\begin{alignat}{10}\label{eq:H}
  &H_{el}=E_C \ket{\phi_c}\bra{\phi_c}&&+\ebare' \sum_\alpha \ket{\psi_\alpha}\bra{\psi_\alpha} + \sum_{\alpha\beta}\HIB_{\alpha\beta}\ket{\psi_\alpha}\bra{\psi_\beta}\nonumber\\
  &E_{ph}=\sum_i \frac{1}{2}\omega_i^2 Q_i^2 \\\nonumber
  &H_{el-ph}=\sum_i Q_i &&\big[\sum_{\alpha\beta}A^i_{\alpha\beta}\ket{\psi_\alpha}\bra{\psi_\beta} \\\nonumber
  &&&+ \sum_\alpha B^i_\alpha (\ket{\psi_\alpha}\bra{\phi_c}+\ket{\phi_c}\bra{\psi_\alpha})\big],
\end{alignat}
where $E_C$ is the energy of the conduction band edge, $\ebare'$ is the energy of the empty isolated-impurity state, the IB Hamiltonian $\HIB$ is zero in the dilute-impurity limit, $A^i$ is responsible for the shift in the equilibrium phonon coordinate when an impurity state is occupied, and $B^i$ causes the transition between the conduction and impurity states. In what follows, we will neglect $B^i$ in describing state energies, as it is important only when IB and CB states are nearly degenerate. The phonon mass has been incorporated into the phonon coordinates (lattice displacements) $Q_i$. The phonon modes include all standard crystalline extended modes in addition to any local vibrational modes around the impurity. 

We begin by considering the standard case of only one impurity, $N=1$, so $\HIB=0$. If the electron is in the CB level, then the energy of the system is $E^C(\vec{Q})=E_C+\sum_i\omega_i^2 Q_i^2/2$. If the electron is in the impurity level, then the energy is
\begin{align}
  E^I(\vec{Q})=\ebare' + \sum_i\frac{1}{2}\omega_i^2 (Q_i+A^i/\omega_i^2)^2 - {A^i}^2/2\omega_i^2. \nonumber
\end{align}
We see that the equilibrium phonon position is $\delta Q_i=-A^i/\omega_i^2$. We define the relaxation energy $\lambda\equiv \sum {A^i}^2/2\omega_i^2$. When the impurity state is filled, its relaxed energy is $\ebare=\ebare'-\lambda$, as shown in Fig.\ \ref{fig:isolated}. We also define $\DEp=E_C-\ebare'$ and $\DE=E_C-\ebare$, where primes indicate empty IB levels. It is customary to define the dimensionless Huang-Rhys factor for each phonon mode as
\begin{align}\label{eq:S}
  S_i\equiv\omega_i\delta Q_i^2/2\hbar={A^i}^2/2\hbar\omega_i^3,
\end{align}
where $\hbar$ is Planck's constant. The total Huang-Rhys factor is $S=\sum_i S_i$. Then $\lambda=\hbar\sum_i S_i \omega_i$. In the case of polar electron-phonon coupling, $S\propto 1/a_B^*$ where $a_B^*$ is the effective Bohr radius of the hydrogenlike impurity wavefunction \cite{Soltani95,Gurskii99,Malm71}. As the defect state becomes more delocalised, $S$ (and thus $\lambda$) decreases. In nonpolar semiconductors, the relationship between $S$ and wavefunction size is, to our knowledge, unexplored. Since the argument for lifetime recovery relies on this relationship, we will assume that it applies.

The energies $E^C$, $E^I$ are equal when the phonon coordinate is such that $\vec Q \cdot \vec A=\DEp$. The activation energy to reach this configuration is $E_a=\sum_i \omega_i^2 Q_i^2/2$. The phonon coordinate $\vec{Q}$ satisfying $E^C(\vec Q)=E^I(\vec Q)$ and minimizing $E_a$ is denoted $\vec Q_c$.  A simple application of Lagrange multipliers shows that 
\begin{align}
  Q_{c,i}=\frac{A^i}{\omega_i^2}\frac{\DEp}{2\lambda},
\end{align}
which gives
\begin{align}\label{eq:Ea single}
  E_a=\frac{1}{2}\sum_i \omega_i^2 Q_{c,i}^2 =\frac{(\DEp)^2}{4\lambda}=\frac{(\DE-\lambda)^2}{4\lambda}.
\end{align}
This is the same activation energy as in Marcus theory, in whose terminology the IB trapping problem is generally in the inverted regime \cite{Marcus56}.

In the high-temperature activated regime, where $k_B T \gg \hbar \omega/2$, the multiphonon trapping process occurs primarily through $\vec Q=\vec Q_c$ \cite{Landsberg91,Huang50}, giving the multiphonon trapping rate
\begin{align}\label{eq:gamma}
  \gamma=\abs{V(\vec{Q}_c)}^2\left(\frac{\pi}{k_B T \lambda}\right)^{1/2} \exp\left(-\frac{E_a}{k_B T}\right),
\end{align}
where $T$ is the temperature, $k_B$ is Boltzmann's constant, and the off-diagonal matrix element $V(\vec{Q}_c)=\vec Q_c \cdot \vec B$ \cite{*[] [{ ch.\ 6}] Landsberg91}.\footnote{For applications such as SRH statistics, the impurity trapping cross section is $\sigma_c=\Omega \gamma/ v_{th}$, where $\Omega$ is the system size and $v_{th}$ is the thermal electron velocity \cite{Landsberg91}. Note that  $B\propto1/\sqrt{\Omega}$, so $\sigma_c$ is independent of system size.} The case of the low temperature tunneling regime will be discussed at the end of this article.

\subsection{Intermediate Band}

We see from Eqs.\ \ref{eq:Ea single}, \ref{eq:gamma} that $\gamma$ increases with $\lambda$ until $\lambda$ is near $\DE$. Thus, one can potentially suppress $\gamma$ by decreasing $\lambda$. For large impurity concentration $N_t$, the impurity states can undergo an insulator-to-metal transition (IMT), with their eigenstates transitioning from localised to extended. Since $\lambda$ decreases with impurity wavefunction size, Luque et al.\ suggested that the IMT would suppress the trapping rates and thus the overall recombination rate $U$ \cite{Luque06}. We show that the extrapolation from the isolated-impurity problem to the many-impurity problem is more complicated. Even when the IB has extended eigenstates, the lattice distortions associated with multiphonon recombination localise electronic states sufficiently to cause multiphonon recombination. The delocalised states in fact can only \emph{increase} the trapping rate.
A related extension to many intermediate states was analysed in Ref.\ \onlinecite{Landsberg88}. Let $\Delta=\max\{\abs{\ehi-\ebare'},\abs{\elo-\ebare'}\}$ be the largest energy shift due to the many-impurity physics. Then $\Delta/\DEp$ is our perturbation parameter, assumed to be small.

\begin{figure}
  \includegraphics[width=\figwidth]{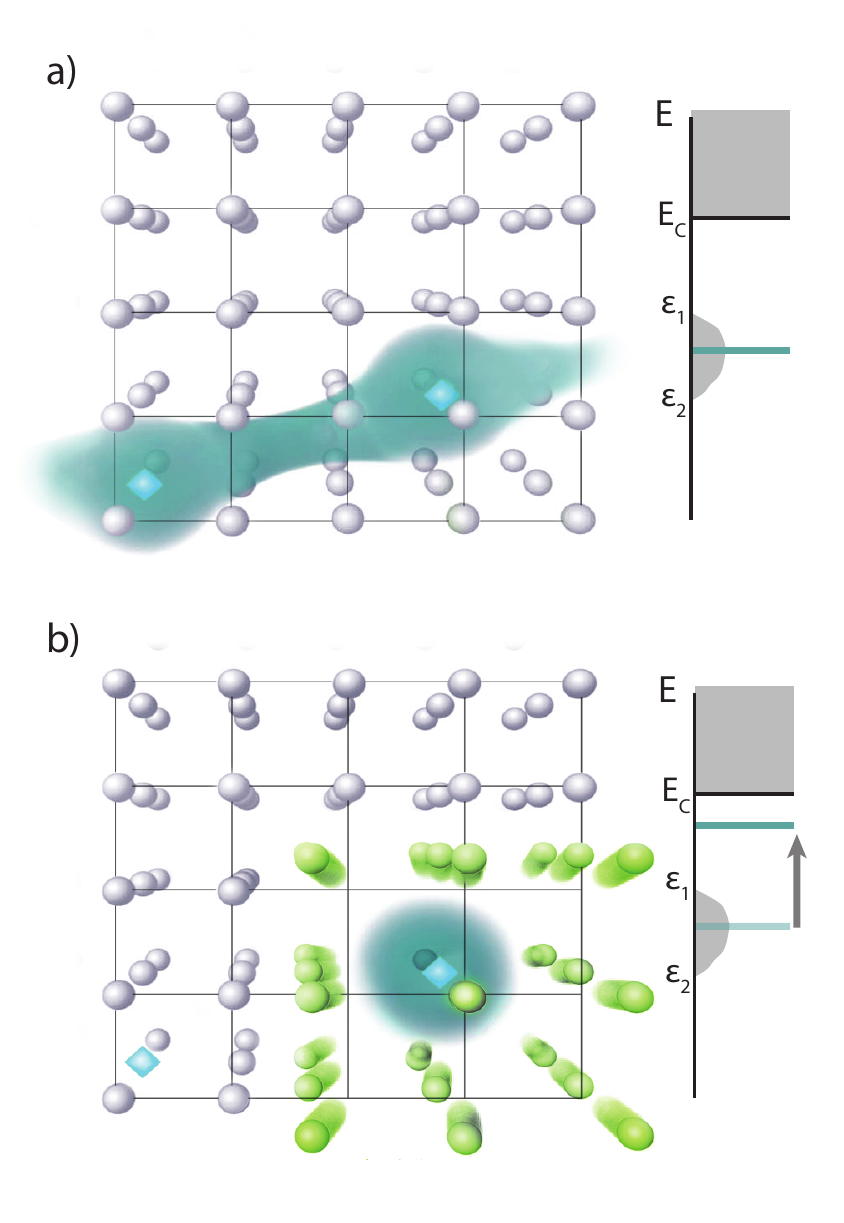}
  \caption{\label{fig:localization} Cartoon of a crystal of gray spherical atoms with blue diamond impurities. a) With the intermediate band unoccupied and the atoms in their ground state positions, an eigenstate in the IB (light green) may be extended across the impurity sites. Its energy is indicated by the green line in the density of states (DOS), at right, which shows the conduction band and the intermediate band. b) When a thermal fluctuation moves the atoms surrounding an impurity away from their equilibrium positions, an eigenstate becomes localised at that impurity; its energy is shown by the green line in the DOS. This occurs because for sufficiently large deviations of the host atoms, the electron-phonon coupling energy is larger than the IB bandwidth, so the eigenstates of the electron-phonon coupling become approximate eigenstates of the IB. When the energy of this localised state equals that of the CB, rapid trapping occurs.}
\end{figure}

We start with the intuitive argument, illustrated in Fig.\ \ref{fig:localization}. The multiphonon trapping process occurs with the lattice in a configuration $\vec Q_c$ where a CB and IB state are degenerate. In a system where $\Delta\ll\ \DEp$, the energy to raise an IB state above the conduction band minimum must come predominantly from the electron-phonon coupling. In such a lattice configuration, the eigenstates of the electron-phonon coupling (i.e., the localised states) are the approximate system eigenstates, with $\HIB$ acting only as a perturbation, regardless of whether the eigenstates of $\HIB$ are localised or extended. Only when $\Delta$ is of similar order to $\DE$ is it possible for $\HIB$ to produce extended states that are still extended when $\vec Q=\vec Q_c$. In order to realise the Luque et al.\ proposal for lifetime recovery \cite{Luque06}, the IB must approach this large-bandwidth limit, having extended states and sufficient bandwidth to resist their localization. However, when $\Delta$ approaches $\DEp$, the IB crosses with the CB, and the material is no longer useful for IBPV. 

Both SRH statistics and the dependence of $\gamma$ (Eq.\ \ref{eq:gamma}) on ionization energy $\DEp$  produce the result that recombination is fastest through states at the centre of the band.  Thus, a perturbation that moves IB energy levels uniformly away from the centre of the band will decrease the recombination rate. This trivial form of lifetime recovery is possible, but a suppression of recombination due to delocalization arising from an insulator-to-metal transition is not.

We develop this argument formally using Eq.\ \ref{eq:H} in the case with $N$ impurity levels. The $A^i$ are now matrices that couple IB states. 
In the basis of localised IB states, the $A^i_{\alpha \beta}$ should be local matrices, either diagonal or with small off-diagonal components, as the phonons mainly act to raise and lower the energy of occupying a given localised state; off-diagonal terms in this basis encourage delocalization, which is not expected for phonons. The calculation below is simplest if we assume the $A^i$ are all diagonal in the localised basis (or otherwise that they all commute), but we will keep their off-diagonal components and formally treat them as small.

The trapping rate is determined by the minimal activation energy $E_a=\sum_i \omega_i^2 Q_i^2/2$ such that an IB state is degenerate with the CB, where we neglect $B^i$. That is, we want to find a vector $\vec{Q}$ such that the highest eigenvalue of the Hamiltonian
\begin{align}
  H_{tot}(\vec{Q})=\HIB+\vec{Q} \cdot \vec{A}
\end{align}
is equal to $\DEp$, and $E_a$ is minimal. Let the highest eigenvalue of $H_{tot}(\vec{Q})$ be $E_0(\vec{Q})$.

\subsubsection{Variational argument}
We now present a variational argument that off-diagonal terms of $\HIB$ (i.e., those that can produce metallic, extended states) always increase the trapping rate. We consider an IB in which the bandwidth is less than $\DEp$, so the IB does not overlap with the CB when $\vec{Q}=0$.
With $\HIB=0$, let $\vec{Q}_0$ be the phonon coordinate such that $E_0^{\HIB=0}(\vec{Q}_0)=\DEp$ and the activation energy, $E_{a}^{\HIB=0}$, is minimal.\footnote{If there are $N$ identical defects in identical environments, then there are $N$ such vectors, one for each localised defect state. Here we consider just one of them.}  Let $\ket{\psi_0}$ be the associated eigenstate; $\ket{\psi_0}$ is localised since $\HIB=0$ and trapping is fastest through localised states.  Assume that $\bra{\psi_0}\HIB\ket{\psi_0}=0$, to eliminate the trivial change of trapping rates due to shifts in the localised states' energies. By the variational principle, $E_0(\vec{Q}_0)\geq \DEp$, since $\bra{\psi_0}H_{tot}\ket{\psi_0}=\DEp$. We assume that $E_0(\vec{Q})$ is a smooth function of $\vec{Q}$. Then since by assumption $E_0(\vec{Q}=0)<\DEp$, there is a $\kappa\leq1$ such that $E_0(\kappa \vec{Q}_0)=\DEp$. The activation energy for this state is $E_a=\kappa^2 E_{a}^{\HIB=0}$. Thus, the trapping rate is greater than or equal to the trapping rate without $\HIB$ at all.

\subsubsection{Perturbative calculation}

This variational result is non-constructive. To understand its physical origin, we perform a perturbative calculation in $\HIB$ to find the activation energy as delocalizing effects are included. This will also allow us to make a prediction for trapping rate changes, including the trivial effect.

First consider $\HIB=0$. As in the variational argument, there is a vector $\vec{Q}_0$ such that the largest eigenvalue of $H_0=\vec{Q}_0 \cdot \vec{A}$ is $\DEp$ and $\sum_i \omega_i^2Q_{0,i}^2$ is minimal. Choose the basis $\{\psi_\alpha\}$ for $\alpha=0\dots N-1$ to be the eigenstates of $H_0$, with eigenvalues $\{E_\alpha\}$ with $E_0=\DEp$.  
Using the constraint $\DEp=\bra{\psi_0}H_0\ket{\psi_0}$, we find $\vec{Q}_0$ using a Lagrange multiplier, minimizing
\begin{align}
  f=\frac{1}{2}\sum_i \omega_i^2 Q_{0,i}^2 +l(\DEp -\vec{Q}_0\cdot \vec{A}_{00}),
\end{align}
where we use the notation $A_{\alpha\beta}\equiv\bra{\psi_\alpha}A\ket{\psi_\beta}$, giving
\begin{align}
  Q_{0,i}&=\frac{\DEp}{2\lambda_0}\frac{A^i_{00}}{\omega_i^2},\label{eq:Q_0i}
\end{align}
and we define the isolated impurity relaxation energy
\begin{align}
  \lambda_0=\sum_i \frac{(A^i_{00})^2}{2\omega_i^2}.\label{eq:lambda0}
\end{align}
Together, these give the activation energy
\begin{align}\label{eq:Ea isolated}
  E_{a}^{0}=\frac{1}{2}\sum_i\omega_i^2 Q_{0,i}^2=\frac{(\DEp)^2}{4\lambda_0},
\end{align}
just as in Eq.\ \ref{eq:Ea single}.

\subsubsection{Including the IB Hamiltonian $\HIB$}

When we include $\HIB$, we must also consider a shift in the phonon vector,
\begin{align}
  \vec{Q}=\vec{Q}_0+\vec{\dQ}.
\end{align}
We consider the perturbation Hamiltonian
\begin{align}
  H_1=\HIB+\vec{\dQ}\cdot \vec{A},
\end{align}
and we seek the self-consistent solution $\vec{\dQ}$ such that the highest eigenvalue of $H_0+H_1$ equals $\DEp$ and $\sum_i \omega_i^2Q_i^2/2$ is minimal.

We consider second-order perturbation theory. Near the configuration $\vec Q_0$, the electron-phonon coupling most strongly perturbs a single localised state, so we can assume that no $E_{\alpha>0}$ is close to degenerate with $E_0$. When $H_1$ is added, the highest eigenvalue is
\begin{align}
  \tilde E_0=E_0+\bra{\psi_0}H_1\ket{\psi_0} + \sum_{\alpha>0} \frac{\abs{\bra{\psi_\alpha} H_1\ket{\psi_0}}^2}{E_0-E_\alpha} + \text{third order}.
\end{align}
We impose the constraint that $\tilde E_0=E_0=\DEp$ by minimizing the function
\begin{align}
  f=&\sum_i \frac{1}{2}\omega_i^2(Q_{0,i}+\dQ_i)^2 + 
  l(\tilde E_0 - E_0),
\end{align}
\begin{align}
\label{eq:del_f}
  \frac{\del f}{\del \dQ_i}=0=&\omega_i^2(Q_{0,i}+\dQ_i) \\\nonumber
  &+ l\left[ A^i_{00} + 2\sum_{\alpha>0}\frac{\HIB_{\alpha0}+\vec{\dQ}\cdot \vec{A}_{\alpha0}}{E_0-E_\alpha}A^i_{\alpha0}\right],
\end{align}
where we have assumed that all the wavefunctions and operators are real.

If the matrices $A^i$ all commute, then $A^i_{\alpha0}\propto \delta_{\alpha0}$. We instead assume that the $A^i_{\alpha0}$ are small for $\alpha>0$ and consider $\HIB$ and $\sum_{\alpha>0} \vec{Q}_c\cdot \vec{A}_{\alpha0}$ to be of the same order.  Then the term $\vec{\dQ}\cdot \vec{A}_{\alpha0} A^i_{\alpha0}$ in Eq.\ \ref{eq:del_f} is third order and can be neglected.
 Note that since the $\ket{\psi_\alpha}$ are eigenstates of $\vec{Q}_0\cdot \vec{A}$,
\begin{align}\label{eq:Q A sum}
  \vec{Q_0}\cdot \vec{A_{\alpha0}}=\delta_{\alpha0}\DEp.
\end{align}
Using this result and Eq.\ \ref{eq:Q_0i},
\begin{align}
  \sum_i A^i_{\alpha0}\frac{A^i_{00}}{\omega_i^2} = \sum_i A^i_{\alpha0} Q_{0,i} \frac{2\lambda_0}{\DEp}=2\lambda_0 \delta_{\alpha0}.
\end{align}
Then we find
\begin{align}
  l&=\frac{-\DEp + \HIB_{00} + \sum_{\alpha>0}\frac{\abs{\HIB_{\alpha0}}^2}{E_0-E_\alpha}}{2\lambda_0} \quad \text{and}\\
  \dQ_i&= -\left(\HIB_{00} + \sum_{\alpha>0}\frac{\abs{\HIB_{\alpha0}}^2}{E_0-E_\alpha}\right) \frac{Q_{0,i}}{\DEp} - \frac{\DEp}{\lambda_0 \omega_i^2}\sum_{\alpha>0}\frac{\HIB_{\alpha0}A^i_{\alpha0}}{E_0-E_\alpha}.\nonumber
\end{align}
Note that the first term of $\vec{\dQ}$ is parallel to $\vec Q_0$. The last term of $\vec{\dQ}$ shifts the direction of the phonon coordinates and is second order (or zero if the $A^i$ commute).

We use this result and Eq.\ \ref{eq:Q_0i} to find the new activation energy
\begin{align}\label{eq:Ea}
  E_a=&E_a^0\left[\left(1-\frac{\HIB_{00}}{\DEp}\right)^2-\frac{2}{\DEp}\sum_{\alpha>0} \frac{\abs{\HIB_{\alpha0}}^2}{E_0-E_\alpha}\right],
\end{align}
where we have kept terms to second order and again have used Eq.\ \ref{eq:Q A sum}. Note that the shift in the direction of the lattice distortion has no effect on $E_a$, at least to second order.

The second order correction to $E_a$ (the second term in Eq.\ \ref{eq:Ea}) is always negative, since $E_0>E_\alpha$ for all $\alpha>0$, thus increasing $\gamma$. These are the terms resulting from the off-diagonal components of $\HIB$, which the variational argument above showed always increase the trapping rate. The first order term represents a trivial change in the trapping rate due to a shift in the energy of the localised states.

We can use this result to find an upper bound on $E_a$ (and associated lower bound on $\gamma$), which is valid through second order. The first order correction simply takes the activation energy for the localised states, Eq.\ \ref{eq:Ea isolated}, and shifts the energy gap $\DEp$ by $\HIB_{00}$.
We consider $\HIB_{00}$ to be unknown, though it could be easily determined within a model for the IB. For example, a standard tight-binding model for the IB would have $\HIB_{00}=0$. Since we consider the bounds of $\HIB$ to be experimentally determinable, we know $\HIB_{00}\geq\elo$. Then we can bound, to second order,
\begin{align}\label{eq:Ea bound}
  E_a\leq \frac{(E_C-\elo)^2}{4\lambda_0}.
\end{align}

If there are $N$ defects (equivalently, $N$ distinct IB states), then each one has an associated $\vec{Q}_c$, whose influence on trapping we can estimate in this same way. Then Eq.\ \ref{eq:Ea bound} gives a lower bound on the overall trapping rate for CB electrons that scales with $N$, just as for independent defects.

We now consider the case of an IB that is not initially empty. If the IB bandwidth is less than $\lambda_0$, then small-polaron formation \cite{Holstein59} will localise the filled IB states, splitting the IB into a lower-energy, filled band and a higher-energy empty IB, which should behave as described in this article.
If the IB bandwidth is significantly greater than $\lambda_0$ and if the IB begins partially occupied, the effective gap to the CB, $\DEp$, of Eq.\ \ref{eq:Ea isolated} will be increased if the localised state $\ket{\psi_0}$ begins partially occupied. An energy cost of up to $\lambda_0$ must be paid to empty the state, which will decrease the trapping rate.

While we have discussed dopant-produced IB's, a similar result holds for highly-mismatched alloys (HMA), in which the IB is formed by the repulsion of the CB from resonances above the CB minimum \cite{Lopez11,Shan99,Walukiewicz00}. In this case, recombination can still proceed through the IB, but there is no isolated impurity state for comparison. Multiphonon trapping will still occur through the localised states in the IB, since they are the eigenstates of the electron-phonon coupling. For any IB, the reorganization energy $\lambda_0$ is determined by the smallest wavefunction that can be made by linear combinations of the states of the IB.\footnote{In the case of dopant-produced IB's, the isolated-dopant wavefunction is presumably the smallest wavefunction in the span of the IB states.} In addition to this $\lambda_0$, the HMA IB can still be characterised by a maximum and minimum energy $\ehi$ and $\elo$ and bandwidth $J=\ehi-\elo$. As long as $J\ll E_C-\ehi$, the localised wavefunctions will become approximate eigenstates of the IB as the lattice is distorted, and multiphonon trapping through the localised states will occur, as described above for the dopant-produced IB. Since the IB in the HMA case forms from the undoped CB, $\lambda_0$ may be considerably smaller than in the deep-level-dopant case, giving a small multiphonon recombination rate; lifetime recovery is still not expected.

The high-temperature limit used in Eq.\ \ref{eq:gamma} is only applicable for $k_B T \gg \hbar \av{\omega}$, where $\av{\omega}$ is the typical phonon frequency. At low temperatures, the phonon momentum must be treated correctly, and tunneling is required to realise the multiphonon trapping process. In the low-temperature tunneling regime
\begin{align}
  \gamma\propto \exp\left\{-\frac{\DE}{\hbar \omega_M} \left[\log\left(\frac{\DE}{\lambda}\right)-1\right]\right\}
\end{align}
where $\omega_M$ is the maximum phonon frequency \cite{Englman70}. The effect of $\HIB$ is effectively to change $\DE$ without changing $\lambda$ (see Eq.\ \ref{eq:S}). With a deep IB, $\DE\gg\lambda$ so $\DE\approx\sqrt{4\lambda E_a}$. In this case, $\gamma$ still decreases exponentially with $\sqrt{E_a}$, and the above analysis of the qualitative effect of delocalised states is still valid.

\section{Valence band trapping}
A similar estimate can be made for the IB to VB trapping process: trapping occurs through the phonon coordinate $\vec{Q}_v$ such that the lowest energy filled IB state is degenerate with a VB state. The relevant Hamiltonian is the same as in Eq.\ \ref{eq:H}, with the addition of a term  $E_V\ket{\phi_v}\bra{\phi_v}$, with $\ket{\phi_v}$ a VB state, where $E_V=E_C-E_g$. We define the energy gap between the filled impurity state and the VB to be $\DE_v=\ebare-E_V$. Then the activation energy to first order is
\begin{align}\label{eq:Ea_v bound}
  E_{a,v}=\frac{(\DE_v-\lambda_0-\HIB_{00})^2}{4\lambda_0}.
\end{align}
Note that the unperturbed phonon coordinate, as in Eq.\ \ref{eq:Q_0i}, is
\begin{align}
  \vec{Q}_{0,v}=-\frac{\DEp_v}{2\lambda_0}\frac{\vec{A}_{00}}{\omega_i^2}=-\frac{\DEp_v}{\DEp}\vec{Q}_0,
\end{align}
where $\DEp_v=\DE_v+\lambda_0$. We see that $\vec{Q}_{0,v}$ is a displacement of the opposite sign from $\vec{Q}_0$ but in the same direction.

Note that if $\HIB_{00}$ is close to $\elo$, as in Eq.\ \ref{eq:Ea bound}, the CB trapping rate is slowed, but the VB trapping rate is increased, and vice versa if $\HIB_{00}$ is close to $\ehi$. The overall recombination rate will be minimised by slowing whichever of the CB and VB trapping rates is rate-limiting. For a functioning IB photovoltaic, both trapping rates must independently be small, so the material can sustain separate quasi-Fermi levels for the CB, IB, and VB populations \cite{Luque97}, as in Fig.\ \ref{fig:IBdevice}.

\section{Prospects for IBPV}
Though nontrivial lifetime recovery in an IB system appears impossible, there is still good reason to think that IB devices can improve PV efficiencies. Consider an IBPV device as illustrated in Fig.\  \ref{fig:IBdevice}. The high-energy photons will be absorbed by the p-type region before reaching the IB layer. Let $\alpha$ be the mean absorptivity of the IB region over the subgap portion of the spectrum. The IB layer width $w$ must be sufficiently wide to absorb a large fraction of subgap photons, so it is desirable to choose $w=c/\alpha$ for $c\approx2-3$. For the IBPV device to outperform the single-junction device with no IB, the IB region must add more current by absorption of subgap photons than it subtracts by enhanced nonradiative recombination, in addition to not significantly changing the voltage.

Let $t$ be the transit time for a CB electron to move from the p-type to n-type side of the device and let $\tau$ be the nonradiative lifetime of CB electrons in the IB region. The IBPV device will have higher efficiency than the single-junction device as long as $\nu\equiv \tau/t\gg1$. If we assume that the built-in voltage $V_{bi}$ is dropped mostly across the IB region, then
\begin{align}
  t=\frac{w^2}{\mu V_{bi}}
\end{align}
where $\mu$ is the electron mobility in the IB region. Thus, for an IB region just thick enough to absorb the subgap light \cite{Guettler70},
\begin{align}
  \nu=\frac{1}{c^2}V_{bi} \mu \alpha^2 \tau.
\end{align}
We roughly expect $\tau\propto 1/N_t$ and $\alpha\propto N_t$ in the IB region. The mobility also declines as $N_t^{-\beta}$, but generally with a small exponent $\beta$ \cite{Arora82,Klaassen92}. Thus, we have reason to believe that high dopant concentrations and strong subgap absorptivities \cite{Crouch04,Olea11,Bob10,Pan11} can produce useful IBPV devices. Full analysis balancing recombination and current generation as in Ref.\ \onlinecite{Keevers94} is required.

Future research efforts should focus on producing highly absorptive, thin layers of IB materials made of dopants which have as-small-as-possible contributions to nonradiative recombination, not on lifetime recovery. Non-dopant-based IBPV proposals, including HMA's \cite{Lopez11} and crystalline systems \cite{Palacios08} may have inherently small trapping rates, so are also promising.

\acknowledgments
We acknowledge helpful comments on the manuscript from Mark Winkler and conversations with Daniel Recht, graphics help from Lauren Kaye, and support from NSF Grant Nos.\ DMR-0934480 and DMR-0906475 and the Harvard University Center for the Environment.
\bibliography{b-Si}
\end{document}